\documentclass[11pt, oneside]{amsart}

\usepackage{amsmath}
\usepackage{amssymb}

\usepackage{color}
\usepackage{dcolumn}
\usepackage{float}
\usepackage{graphicx}
\usepackage[latin9]{inputenc}
\usepackage{multirow}
\usepackage{rotating}
\usepackage{subfigure} 
\usepackage{psfrag}
\usepackage{tabularx}
\usepackage[hyphens]{url}
\usepackage{wrapfig}

\usepackage[bookmarks, bookmarksopen, bookmarksnumbered]{hyperref}
\usepackage[all]{hypcap}
\definecolor{orange}{rgb}{1.0,0.3,0.0}
\definecolor{violet}{rgb}{0.75,0,1}
\definecolor{darkgreen}{rgb}{0,0.6,0}
\definecolor{cyan}{rgb}{0.2,0.7,0.7}
\definecolor{blueish}{rgb}{0.2,0.2,0.8}

\usepackage[top=2.5cm, bottom=2.5cm, left=2.5cm, right=2.5cm]{geometry}

\sloppypar

\begin{document}

\title[]{First Workshop on Sustainable Software for Science: Practice and Experiences (WSSSPE): Submission and Peer-Review Process, and Results}

\author{Daniel S. Katz$^{(1)}$, Gabrielle Allen$^{(2)}$, Neil Chue Hong$^{(3)}$, \\
Manish Parashar$^{(4)}$, and David Proctor$^{(1)}$ }

\thanks{{}$^{(1)}$ National Science Foundation, Arlington, VA, USA}

\thanks{{}$^{(2)}$ Skolkovo Institute of Science and Technology, Moscow, Russian Federation}

\thanks{{}$^{(3)}$ Software Sustainability Institute, University of Edinburgh, Edinburgh, UK}
  
\thanks{{}$^{(4)}$ Rutgers Discovery Informatics Institute, Rutgers University, New Brunswick, NJ, USA}

\begin{abstract}
This technical report discusses the submission and peer-review process used by the First Workshop on on Sustainable Software for Science: Practice and Experiences (WSSSPE) and the results of that process.  It is intended to record both this alternative model as well as the papers associated with the workshop that resulted from that process.
\end{abstract}

\maketitle

\section{Introduction}

%
%
%
%

The First Workshop on on Sustainable Software for Science: Practice and Experiences (WSSSPE)\footnote{\url{http://wssspe.researchcomputing.org.uk}} will be held on Sunday, 17 November 2013, in conjunction with the 2013 International Conference for High Performance Computing, Networking, Storage and Analysis (SC13)\footnote{\url{http://sc13.supercomputing.org}}.

Progress in scientific research is dependent on the quality and accessibility of software at all levels and it is now critical to address many challenges related to the development, deployment, and maintenance of reusable software. In addition, it is essential that scientists, researchers, and students are able to learn and adopt software-related skills and methodologies. Established researchers are already acquiring some of these skills, and in particular a specialized class of software developers is emerging in academic environments who are an integral and embedded part of successful research teams. The WSSSPE workshop was intended to provide a forum for discussion of the challenges, including both positions and experiences. The short papers and discussion were archived to provide a basis for continued discussion, and the workshop was intended to feed into the collaborative writing of one or more journal publications.

\section{Submissions}

The workshop call for paper included:

\begin{quote}
In practice, scientific software activities are part of an ecosystem where key roles are held by developers, users, and funders.  All three groups supply resources to the ecosystem, as well as requirements that bound it.  Roughly following the example of NSF's Vision and Strategy for Software (\url{http://www.nsf.gov/publications/pub_summ.jsp?ods_key=nsf12113})~\cite{NSF_software_vision}, the ecosystem may be viewed as having challenges related to:

\begin{itemize}
\item the development process that leads to new software
\begin{itemize}
\item how fundamental research in computer science or science/engineering domains is turned  into reusable software
\item software created as a by-product of research
\item impact of computer science research on the development of scientific software
\end{itemize}
\item the support and maintenance of existing software, including software engineering
\begin{itemize}
\item governance, business, and sustainability models
\item the role of community software repositories, their operation and sustainability
\end{itemize}
\item the role of open source communities or industry
\item use of the software
\begin{itemize}
\item growing communities
\item reproducibility, transparency needs that may be unique to science
\end{itemize}
\item policy issues, such as
\begin{itemize}
\item measuring usage and impact
\item software credit, attribution, incentive, and reward
\item career paths for developers and institutional roles
\item issues related to multiple organizations and multiple countries, such as intellectual property, licensing, etc.
\item mechanisms and venues for publishing software, and the role of publishers
\end{itemize}
\item education and training
\end{itemize}

\end{quote}

Based on the goal of encouraging a wide range of submissions from those involved in software practice, ranging from initial thoughts and partial studies to mature deployments, the organizers wanted to make submission as easy as possible.  The call for papers stated:

\begin{quote}
We invite short (4-page) position/experience reports that will be used to organize panel and discussion sessions.  These papers will be archived by a third-party service, and provided DOIs. We encourage submitters to license their papers under a Creative Commons license that encourages sharing and remixing, as we will combine ideas (with attribution) into the outcomes of the workshop.  An interactive site will be created to link these papers and the workshop discussion, with options for later comments and contributions.  Contributions will be peer-reviewed for relevance and originality before the links are added to the workshop site; contributions will also be used to determine discussion topics and panelists.  We will also plan one or more papers to be collaboratively developed by the contributors, based on the panels and discussions.
\end{quote}

58 submissions were received, and almost all submitters used either arXiv\footnote{\url{http://arxiv.org}} or figshare\footnote{\url{http://figshare.com}} to self-publish their papers.  

\section{Peer-Review}

A peer review process followed the submissions, where the 58 papers received 181 reviews, an average of 3.12 reviews per paper.  Reviews were completed using a Google form, which allowed reviewers to provide scores on relevance and comments to the organizers, which were used to decide which papers to associate with the workshop, and comments to the authors, which were provided back to the authors to allow them to improve their papers.

The organizers decided to list 54 of the papers as significantly contributing to the workshop, a very high acceptance rate, but one that is reasonable, given the goal of broad participation and the fact that the reports were already self-published.  The papers were also grouped into 3 areas, each of which will be associated with a panel and discussion at the workshop.

\section{Results}

The contributed papers that will be discussed at the workshop follow, listed by area.

\subsection{Developing, Deploying and Supporting Software}

\subsubsection{Development Experiences}

\begin{itemize}

\item Mark C. Miller, Lori Diachin, Satish Balay, Lois Curfman McInnes, Barry Smith. Package Management Practices Essential for Interoperability: Lessons Learned and Strategies Developed for FASTMath \cite{Miller_WSSSPE}

\item Karl W. Broman, Thirteen years of R/qtl: Just barely sustainable \cite{Broman_WSSSPE}

\item Charles R. Ferenbaugh, Experiments in Sustainable Software Practices for Future Architectures \cite{Ferenbaugh_WSSSPE}

\item Eric G Stephan, Todd O Elsethagen, Kerstin Kleese van Dam, Laura Riihimaki. What Comes First, the OWL or the Bean? \cite{Stephan_WSSSPE}

\item Derek R. Gaston, John Peterson, Cody J. Permann, David Andrs, Andrew E. Slaughter, Jason M. Miller, Continuous Integration for Concurrent Computational Framework and Application Development \cite{Gaston_WSSSPE}

\item Anshu Dubey, B. Van Straalen. Experiences from Software Engineering of Large Scale AMR Multiphysics Code Frameworks \cite{Dubey_WSSSPE}

\item Markus Blatt. DUNE as an Example of Sustainable Open Source Scientific Software Development \cite{Blatt_WSSSPE}

\item David Koop, Juliana Freiere, Cl\'{a}udio T. Silva, Enabling Reproducible Science with VisTrails~\cite{Koop_WSSSPE}

\item Sean Ahern, Eric Brugger, Brad Whitlock, Jeremy S. Meredith, Kathleen Biagas, Mark C. Miller, Hank Childs, VisIt: Experiences with Sustainable Software \cite{Ahern_WSSSPE}

\item Sou-Cheng (Terrya) Choi. MINRES-QLP Pack and Reliable Reproducible Research via Staunch Scientific Software \cite{Choi_WSSSPE}

\item Michael Crusoe, C. Titus Brown. Walking the talk: adopting and adapting sustainable scientific software development processes in a small biology lab \cite{Crusoe_WSSSPE}

\item Dhabaleswar K. Panda, Karen Tomko, Karl Schulz, Amitava Majumdar. The MVAPICH Project: Evolution and Sustainability of an Open Source Production Quality MPI Library for HPC \cite{Panda_WSSSPE}

\item Eric M. Heien, Todd L. Miller, Becky Gietzel, Louise H. Kellogg. Experiences with Automated Build and Test for Geodynamics Simulation Codes \cite{Heien_WSSSPE}

\end{itemize}

\subsubsection{Deployment, Support, and Maintenance of Existing Software}

\begin{itemize}

\item Henri Casanova, Arnaud Giersch, Arnaud Legrand, Martin Quinson, Fr\'{e}d\'{e}ric Suter. SimGrid: a Sustained Effort for the Versatile Simulation of Large Scale Distributed Systems~\cite{Casanova_WSSSPE}

\item Erik Trainer, Chalalai Chaihirunkarn, James Herbsleb. The Big Effects of Short-term Efforts: A Catalyst for Community Engagement in Scientific Software \cite{Trainer_WSSSPE}

%

\item Jeremy Cohen, Chris Cantwell, Neil Chue Hong, David Moxey, Malcolm Illingworth, Andrew Turner, John Darlington, Spencer Sherwin. Simplifying the Development, Use and Sustainability of HPC Software \cite{Cohen_WSSSPE}

\item Jaroslaw Slawinski, Vaidy Sunderam. Towards Semi-Automatic Deployment of Scientific and Engineering Applications \cite{Slawinski_WSSSPE}

\end{itemize}

\subsubsection{Best Practices, Challenges, and Recommendations}

\begin{itemize}

\item Andreas Prli\'{c}, James B. Procter. Ten Simple Rules for the Open Development of Scientific Software \cite{Prlic_WSSSPE}

\item Anshu Dubey, S. Brandt, R. Brower, M. Giles, P. Hovland, D. Q. Lamb, F. L\:{o}ffler, B. Norris, B. O'Shea, C. Rebbi, M. Snir, R. Thakur, Software Abstractions and Methodologies for HPC Simulation Codes on Future Architectures \cite{Dubey2_WSSSPE}

\item Jeffrey Carver, George K. Thiruvathukal. Software Engineering Need Not Be Difficult \cite{Carver_WSSSPE}

\item Craig A. Stewart, Julie Wernert, Eric A. Wernert, William K. Barnett, Von Welch. Initial Findings from a Study of Best Practices and Models for Cyberinfrastructure Software Sustainability \cite{Stewart_WSSSPE}

\item Jed Brown, Matthew Knepley, Barry Smith. Run-time extensibility: anything less is unsustainable \cite{Brown_WSSSPE}

\item Shel Swenson, Yogesh Simmhan, Viktor Prasanna, Manish Parashar, Jason Riedy, David Bader, Richard Vuduc. Sustainable Software Development for Next-Gen Sequencing (NGS) Bioinformatics on Emerging Platforms \cite{Swenson_WSSSPE}
 
\end{itemize}

\subsection{Policy}

\subsubsection{Modeling Sustainability}

\begin{itemize}

\item Coral Calero, M. Angeles Moraga, Manuel F. Bertoa. Towards a Software Product Sustainability Model \cite{Calero_WSSSPE}

\item Colin C. Venters, Lydia Lau, Michael K. Griffiths, Violeta Holmes, Rupert R. Ward, Jie Xu. The Blind Men and the Elephant: Towards a Software Sustainability Architectural Evaluation Framework \cite{Venters_WSSSPE}

\item Marlon Pierce, Suresh Marru, Chris Mattmann. Sustainable Cyberinfrastructure Software Through Open Governance \cite{Pierce_WSSSPE}

\item Daniel S. Katz, David Proctor. A Framework for Discussing e-Research Infrastructure Sustainability \cite{Katz_WSSSPE}

\item Christopher Lenhardt, Stanley Ahalt, Brian Blanton, Laura Christopherson, Ray Idaszak. Data Management Lifecycle and Software Lifecycle Management in the Context of Conducting Science \cite{Lenhardt_WSSSPE}

\item Nicholas Weber, Andrea Thomer, Michael Twidale. Niche Modeling: Ecological Metaphors for Sustainable Software in Science \cite{Weber_WSSSPE}

\end{itemize}

\subsubsection{Credit, Citation, Impact}

\begin{itemize}

\item Matthew Knepley, Jed Brown, Lois Curfman McInnes, Barry Smith. Accurately Citing Software and Algorithms Used in Publications \cite{Knepley_WSSSPE}

\item Jason Priem, Heather Piwowar. Toward a comprehensive impact report for every software project \cite{Priem_WSSSPE}

\item Daniel S. Katz. Citation and Attribution of Digital Products: Social and Technological Concerns \cite{Katz2_WSSSPE}

\item Neil Chue Hong, Brian Hole, Samuel Moore. Software Papers: improving the reusability and sustainability of scientific software \cite{Chue_Hong_WSSSPE}

\end{itemize}

In addition, the following paper from another area will also be discussed in this area.

\begin{itemize}

\item Frank L\"{o}ffler, Steven R. Brandt, Gabrielle Allen and Erik Schnetter. Cactus: Issues for Sustainable Simulation Software \cite{Loffler_WSSSPE}

\end{itemize}

\subsubsection{Reproducibility}

\begin{itemize}

\item Victoria Stodden, Sheila Miguez. Best Practices for Computational Science: Software Infrastructure and Environments for Reproducible and Extensible Research \cite{Stodden_WSSSPE}

\end{itemize}

\subsubsection{Implementing Policy}

\begin{itemize}

\item Randy Heiland, Betsy Thomas, Von Welch, Craig Jackson. Toward a Research Software Security Maturity Model \cite{Heiland_WSSSPE}

\item Brian Blanton, Chris Lenhardt, A User Perspective on Sustainable Scientific Software \cite{Blanton_WSSSPE}

\item Daisie Huang, Hilmar Lapp. Software Engineering as Instrumentation for the Long Tail of Scientific Software \cite{Huang_WSSSPE}

\item Rich Wolski, Chandra Krintz, Hiranya Jayathilaka, Stratos Dimopoulos, Alexander Pucher. Developing Systems for API Governance \cite{Wolski_WSSSPE}
 
\end{itemize}

\subsection{Communities}

\subsubsection{Communities}

\begin{itemize}

\item L. Christopherson, R. Idaszak, S. Ahalt. Developing Scientific Software through the Open Community Engagement Process \cite{Christopherson_WSSSPE}

\item Reagan Moore. Extensible Generic Data Management Software \cite{Moore_WSSSPE}

\item Karen Cranston, Todd Vision, Brian O'Meara, Hilmar Lapp. A grassroots approach to software sustainability \cite{Cranston_WSSSPE}

\item J.-L. Vay, C. G. R. Geddes, A. Koniges, A. Friedman, D. P. Grote, D. L. Bruhwiler. White Paper on DOE-HEP Accelerator Modeling Science Activities \cite{Vay_WSSSPE}

\item Marlon Pierce, Suresh Marru, Mark A. Miller, Amit Majumdar, Borries Demeler. Science Gateway Operational Sustainability: Adopting a Platform-as-a-Service Approach \cite{Pierce2_WSSSPE}

\item Lynn Zentner, Michael Zentner, Victoria Farnsworth, Michael McLennan, Krishna Madhavan, and Gerhard Klimeck, nanoHUB.org: Experiences and Challenges in Software Sustainability for a Large Scientific Community \cite{Zentner_WSSSPE}

\item Andy Terrel. Sustaining the Python Scientific Software Community \cite{Terrel_WSSSPE}

\item Frank L\"{o}ffler, Steven R. Brandt, Gabrielle Allen and Erik Schnetter. Cactus: Issues for Sustainable Simulation Software \cite{Loffler_WSSSPE}

\item Ketan Maheshwari, David Kelly, Scott J. Krieder, Justin M. Wozniak, Daniel S. Katz, Mei Zhi-Gang, Mainak Mookherjee. Reusability in Science: From Initial User Engagement to Dissemination of Results \cite{Maheshwari_WSSSPE}

\item Nancy Wilkins-Diehr, Katherine Lawrence, Linda Hayden, Marlon Pierce, Suresh Marru, Michael McLennan, Michael Zentner, Rion Dooley, Dan Stanzione. Science Gateways and the Importance of Sustainability \cite{Wilkins-Diehr_WSSSPE}

\item Edmund Hart, Carl Boettiger, Karthik Ram, Scott Chamberlain. rOpenSci -- a collaborative effort to develop R-based tools for facilitating Open Science \cite{Hart_WSSSPE}

\end{itemize}

In addition, the following paper from another area will also be discussed in this area.

\begin{itemize}

\item Marcus Hanwell, Amitha Perera, Wes Turner, Patrick O'Leary, Katie Osterdahl, Bill Hoffman, Will Schroeder. Sustainable Software Ecosystems for Open Science \cite{Hanwell_WSSSPE}

\end{itemize}

\subsubsection{Industry \& Economic Models}

\begin{itemize}

\item Anne C. Elster. Software for Science: Some Personal Reflections \cite{Elster_WSSSPE}

\item Ian Foster, Vas Vasiliadis, Steven Tuecke. Software as a Service as a path to software sustainability \cite{Foster_WSSSPE}

\item Marcus Hanwell, Amitha Perera, Wes Turner, Patrick O'Leary, Katie Osterdahl, Bill Hoffman, Will Schroeder. Sustainable Software Ecosystems for Open Science \cite{Hanwell_WSSSPE}

\end{itemize}

In addition, the following papers from other areas will also be discussed in this area.

\begin{itemize}

\item Brian Blanton, Chris Lenhardt, A User Perspective on Sustainable Scientific Software \cite{Blanton_WSSSPE}

\item Markus Blatt. DUNE as an Example of Sustainable Open Source Scientific Software Development \cite{Blatt_WSSSPE}

\item Dhabaleswar K. Panda, Karen Tomko, Karl Schulz, Amitava Majumdar. The MVAPICH Project: Evolution and Sustainability of an Open Source Production Quality MPI Library for HPC \cite{Panda_WSSSPE}

\item Andy Terrel. Sustaining the Python Scientific Software Community \cite{Terrel_WSSSPE}

\end{itemize}

\subsubsection{Education \& Training}

\begin{itemize}

\item Ivan Girotto, Axel Kohlmeyer, David Grellscheid, Shawn T. Brown. Advanced Techniques for Scientific Programming and Collaborative Development of Open Source Software Packages at the International Centre for Theoretical Physics (ICTP) \cite{Girotto_WSSSPE}

\item Thomas Crawford. On the Development of Sustainable Software for Computational Chemistry \cite{Crawford_WSSSPE}

\end{itemize}

In addition, the following papers from other areas will also be discussed in this area.

\begin{itemize}

\item Charles R. Ferenbaugh, Experiments in Sustainable Software Practices for Future Architectures \cite{Ferenbaugh_WSSSPE}

\item David Koop, Juliana Freiere, Cl\'{a}udio T. Silva, Enabling Reproducible Science with VisTrails~\cite{Koop_WSSSPE}

\item Sean Ahern, Eric Brugger, Brad Whitlock, Jeremy S. Meredith, Kathleen Biagas, Mark C. Miller, Hank Childs, VisIt: Experiences with Sustainable Software \cite{Ahern_WSSSPE}

\item Sou-Cheng (Terrya) Choi. MINRES-QLP Pack and Reliable Reproducible Research via Staunch Scientific Software \cite{Choi_WSSSPE}

\item Frank L\"{o}ffler, Steven R. Brandt, Gabrielle Allen and Erik Schnetter. Cactus: Issues for Sustainable Simulation Software \cite{Loffler_WSSSPE}

\item Erik Trainer, Chalalai Chaihirunkarn, James Herbsleb. The Big Effects of Short-term Efforts: A Catalyst for Community Engagement in Scientific Software \cite{Trainer_WSSSPE}

\end{itemize}

\section{Conclusions}

The WSSSPE workshop has begun an experiment in how we can collaboratively build a workshop agenda.  However, contributors also want to get credit for their participation in the process.  And the workshop organizers want to make sure that the workshop content and their efforts are recorded.  Ideally, there would be a service that would be able to index the contributions to the workshop, serving the authors, the organizers, and the larger community.  But since there isn't such a service today, the workshop organizers are writing this initial report and making use of arXiv as a partial solution to provide a record of the workshop.

After the workshop, one or more additional papers will be created that will include the discussions at the workshop.  These papers will likely have many authors, and may be submitted to peer-reviewed journals.

\bibliographystyle{plain}

\bibliography{wssspe}
\end{document}